\documentclass[aps,prb,twocolumn,preprintnumbers, showpacs,superscriptaddress]{revtex4}
\usepackage[final]{graphicx}
\usepackage{dcolumn}
\usepackage{bm}
\usepackage{floatflt,epsfig}
\usepackage{float}
\usepackage{amsmath}
\usepackage{amsfonts}
\usepackage{amssymb}
\usepackage{mathrsfs}
\usepackage[latin1]{inputenc}

\DeclareTextSymbol{\degre}{T1}{6}
\DeclareTextSymbol{\degre}{OT1}{23}

\begin{document}
\title{High shock release in ultrafast laser irradiated metals: \\Scenario for material ejection}% Force line breaks with \\
\author{J.P. Colombier}
\affiliation{CEA/DAM Ile de France, Dept de Physique Th\'eorique et Appliqu\'ee, BP 12, 91680 Bruy\`eres-le-Ch\^{a}tel, France}
\affiliation{Laboratoire Hubert Curien, Universit\'e Jean Monnet, UMR CNRS 5516, 18 rue Benoît Lauras, 42000 Saint-Etienne, France}
\author{P. Combis}
\affiliation{CEA/DAM Ile de France, Dept de Physique Th\'eorique et Appliqu\'ee, BP 12, 91680 Bruy\`eres-le-Ch\^{a}tel, France}
\author{R. Stoian}
\affiliation{Laboratoire Hubert Curien, Universit\'e Jean Monnet, UMR CNRS 5516, 18 rue Benoît Lauras, 42000 Saint-Etienne, France}
\author{E. Audouard}
\affiliation{Laboratoire Hubert Curien, Universit\'e Jean Monnet, UMR CNRS 5516, 18 rue Benoît Lauras, 42000 Saint-Etienne, France}%
\date{\today}

\begin{abstract}

We present one-dimensional numerical simulations describing the behavior of solid matter exposed to subpicosecond near infrared pulsed laser radiation. We point out to the role of strong isochoric heating as a mechanism for producing highly non-equilibrium thermodynamic states. In the case of metals, the conditions of material ejection from the surface are discussed in a hydrodynamic context, allowing correlation of the thermodynamic features with ablation mechanisms. A convenient synthetic representation of the thermodynamic processes is presented, emphasizing different competitive pathways of material ejection. Based on the study of the relaxation and cooling processes which constrain the system to follow original thermodynamic paths, we establish that the metal surface can exhibit several kinds of phase evolution which can result in phase explosion or fragmentation. An estimation of the amount of material exceeding the specific energy required for melting is reported for copper and aluminum and a theoretical value of the limit-size of the recast material after ultrashort laser irradiation is determined. Ablation by mechanical fragmentation is also analysed and compared to experimental data for aluminum subjected to high tensile pressures and ultrafast loading rates. Spallation is expected to occur at the rear surface of the aluminum foils and a comparison with simulation results can determine a spall strength value related to high strain rates.
\end{abstract}


\maketitle
\section{Introduction}
The understanding of matter ablation under pulsed laser irradiation is a long-standing topic of condensed matter physics and materials science.\cite{Anisimov02} The employment of ultrashort laser pulse have opened the possibility to observe laser-induced transformations on ultrafast time scales. To give a complete description of the mechanisms preceding and determining material ejection from surfaces, several physical processes, sometimes occurring simultaneously, have to be examined. The first process involves the energy deposition of the laser pulse and its transfer to the matter constitutive elements. Then, energy distribution and related effects are strongly correlated with the temporal evolution of the transport properties. Depending on the intensity of the pulse, a local mesoscopic transformation occurs on a picosecond time scale, primarily due to a thermodynamic phase transformation.\cite{Rethfeld02} The final process is ablation, which is related to the macroscopic ejection of matter in the solid, liquid, or gas phase. It has been shown recently that the quantity of ejected matter can be numerically estimated.\cite{Colombier05} A more detailed study of solid to plasma transformation, subsequent to numerous intermediate thermodynamic states and associated material deformations are required for a better understanding of the mechanisms involved in the ablation process. In this paper, attention is drawn to the role of shock release on the production of non-equilibrium thermodynamic states and its influence on the ablation mechanisms in the case of selected metals.

The importance of accessing information on transition states should be viewed in a broad context. The understanding of the ultrafast transformations of material is central to both applied and fundamental physics. Theoretical studies present a technological interest in the optimization of material micro-machining and represent a scheme for the understanding of multi-scale processes related to matter transformations. A large amount of work has been devoted to give successive snapshots of the thermodynamic states of matter preceding or following material ejection. For example, time-resolved optical microscopy experiments performed by Sokolowski-Tinten \emph{et al.}\cite{Sokolowski98} on laser-excited materials have revealed the appearance of Newton rings within several tens of picoseconds after irradiation, which have been attributed to the presence of a metastable, inhomogeneous state of matter.\cite{Sokolowski98,Inogamov99a} A manifestation of a rarefaction wave, generated by crossing particular states of the phase diagram, has been also suggested to explain these observations.\cite{Bulgakova99} Based on a van der Waals type equation of state (EOS), simulations performed at thermal equilibrium have shown a spinodal decomposition process occurring near the liquid-vapor critical point.\cite{Vidal01} Optical probing of the properties of the isentropes lying above the critical point has been previously proposed by Celliers \emph{et al.}\cite{Celliers93} to investigate transport properties of such thermodynamic states. More recently, molecular-dynamics calculations have indicated various mechanisms, such as fragmentation, phase explosion,  and vaporization, resulting from different trajectories in the thermodynamic phase-space. Each one is specific to a way of solid ablation in different fluence regimes.\cite{Perez02}

The theoretical problem consists mainly in achieving simulations of the laser-interaction process with sufficient completeness to reproduce the observed dynamics and to predict ways of improvement. In this way, the investigation approach has to contain optical, thermal, hydrodynamical, and mechanical models of non-equilibrium to reproduce subsequent interaction features. Komashko \emph{et al.}\cite{Komashko99} have reported simulation results which correlate laser properties and material removal efficiency, obtained using the one-dimensional radiation hydrodynamic code HYADES enhanced with a wave-equation solver. Similarly, using the hydrodynamic code MULTI that includes electronic heat conduction, Eidmann \emph{et al.} performed solid to plasma simulations of the subpicosecond laser-interaction with aluminum for a wide range of laser intensities.\cite{Eidmann00} Considering the previous works, none of them insist on the evolution of thermodynamic conditions. A more general view, which can be potentially provided by all the
mentioned models, is still lacking, determining a certain loss of details concerning the disintegration mechanisms of the heated metal. The aim of our work is to gather non-equilibrium excitation and transport models in a unified fluid approach and to extend the thermodynamic features, as revealed by the Molecular Dynamics (MD) approach on small scales, to larger volume simulations. In fact, MD possibilities are limited to a superficial region within the irradiation spot and the simulation dimension is restricted to a few hundreds of nanometers in depth.\cite{Perez03} Hydrocodes are more suitable for studying shock wave formation and propagation for systems of mesoscopic scales, and the two approaches are complementary. To investigate the capability of reporting thermodynamic insights within the framework of non-equilibrium fluid models, specific investigations implying strong expansion dynamics have been performed using the 1D hydrocode ESTHER.\cite{Bonneau04,Colombier05} Details about the main assumption and the formalism of this code are provided in the next section. Using this approach, our intention is to shed some light on some of the thermodynamical and mechanical aspects of ultrafast laser ablation.

Due to the energy deposition of ultrashort laser pulse in a swift manner on a thin layer of matter, electrons are heated quasi-isochorically in the solid sample. Actually, the induced deformations are driven by the shock and rarefaction waves which travel at a velocity where a lower estimate is given by the sound velocity. At this velocity, it takes for the shock wave about 10 ps to travel 50 nm in the metal. This time is similar to the electron-ion relaxation time and expansion is insignificant on the timescale of the laser pulse. The heated matter will expand quasi-isentropically, will pass through unconventional physical states, and enter new regions of the phase diagram. In this study, we demonstrate the role of electronic contribution responsible for generating specific paths that lead to material ejection under ultrafast laser irradiation. Once the maximum degree of non-equilibrium is reached, the role of the electronic pressure during relaxation is outlined in order to investigate the sequence of non-conventional thermodynamic states. To extend the analysis of the thermodynamic paths observed during and after interaction with ultrashort pulses, we provide an interpretation of the thermodynamic diagrams which allows an estimation of the amount of metal that undergoes mesoscopic transformation. This estimation could present a direct interest for post-process analysis of ablation in terms of quality and efficiency. Finally, we have also considered mechanical disintegration following a discussion related to spall experiments performed by Tamura \emph{et al}.\cite{Tamura01} Comparison of these results with those obtained in our simulation show a fairly good agreement, suggesting that the low-strain rate value of the spall strength can be applied on short time scales.

The paper is organized as follows. In section II we briefly describe our model of electron-ion non-equilibrium interaction and the hydrodynamic simulation technique. Concentrating on the results of the simulations, we present in section III independent thermodynamic paths followed by different layers inside the material. The results of these simulations are discussed in the context of the possibility of material ejection. In section IV we provide a numerical quantification of the remaining liquid on the solid substrate. In section V we compare profiles of computational stress to current experimental observations of spall formation. In section VI conclusions are drawn.

\section{Electron-ion non-equilibrium}
The metal response under ultrashort pulse laser irradiation has been simulated using various numerical techniques that emphasize particular issues of the matter evolution and include a variety of hydrodynamical processes.\cite{Eidmann00,Laville02,Bulgakova01} To be consistent with the experimentally measured properties of the material under irradiation, suitable models are needed to describe the absorption and the intermediate stages related to phase transformations during the solid to plasma transition. A set of accurate thermophysical parameters has to be inserted in the simulation to provide a correct picture of the excited matter. For a laser pulse of $\tau=150$~fs (FWHM), the wavelength, $\lambda=800$~nm, is smaller than the coherence length $L_{\tau}=c\tau$, where $c$ is the speed of light. Consequently, to determine optical solid and plasma response, Maxwell equations can be reduced to the Helmholtz equation by using slowly-varying-envelope approximation for the electric field. In addition, if the pulse duration is less than a sonic wave period, $\tau<\lambda/c_{S}$, where $c_{S}$ is the sound velocity, we can assume that hydrodynamic processes occur on a time scale longer than the characteristic scale for the evolution of the optical parameters. As a consequence, the resolution of the stationary equations of electrodynamics is sufficient to calculate radiation absorption. The large difference between the Fermi electron velocity and the sound velocity requires distinguishing between the electronic and ionic thermal evolutions during irradiation.\cite{Anisimov74,Kaganov57} The electron-ion  non-equilibrium produces a hydrodynamic decoupling where electronic pressure contributes strongly to the material deformation. General aspects of the code have been described in detail elsewhere,\cite{Colombier05} and the main assumptions were given above to facilitate understanding.

The hydrodynamic code used in this work solves the one-dimensional fluid equations for the conservation of energy, momentum, and mass, in a Lagrangian formalism. The specificity of this approach is the flexible dimension of the cells which deform together with the material and permits an accurate description of shock propagation. Nevertheless, there is no mass transport between the cells which imposes limitation on the description of gas-phase transformation. The target material is treated as a compressible fluid composed of electronic and ionic subsystems with independent internal energy, temperature, and pressure. In addition, the radiation field corresponding to the incident laser pulse is absorbed in the material through the classical inverse bremsstrahlung involving electron-ion collisions. Even with a plane-polarized wave reflecting from a semi-infinite metal, estimation of the amplitude and the phase of the field is not simple and a numerical calculation of the self-consistent response is required. The absorption of the incident electromagnetic wave by the inhomogeneous medium is calculated based on the Helmholtz equation involving a dynamic behavior of complex refractive indices. According to the Drude model, the refractive index evolution depends on the collision time which is initially determined at solid density based on available experimental data.\cite{Palik} During the laser pulse, both the temperature increase and the density decrease affect the laser absorption and optical parameters are calculated using an interpolation between the standard condition data\cite{Palik} and values given by conductivity models of dense plasmas.\cite{Ebeling} From a microscopic point of view, the energy contained in an incident photon is absorbed by a free electron in a Coulombian field, which remains in the conduction zone but is promoted in a state with higher energy. Excited electrons interact with lattice phonons and with other electrons, transferring their energy to the surroundings. During the laser pulse, electrons which have absorbed photons undergo many collisions between themselves but few with lattice phonons. The energy absorbed by an electron is distributed between electrons on a time scale of a few tens of femtoseconds and is then transferred to the lattice in several picoseconds.\cite{Fann92} Our calculations begin with the assumption that the light energy is instantly transformed into heat inside the electron gas. Thus, it is considered that the local equilibrium condition in the electronic system is verified during the entire pulse. Consequently, it is supposed that a certain electronic temperature is defined at each point. In this context, regular equations for the description of the heat flow have been used and the Two Temperature Model (TTM) has been implemented in our fluid treatment.\cite{Anisimov74}  As in Refs. [\onlinecite{Inogamov99a,Inogamov99b}], we have employed wide-range equations of state in tabular form, which have been the subject of extensive development, and for which a great deal of high-quality experimental data is available.~\cite{Bushman93} An additional electronic contribution to the total pressure has been artificially inserted in the equilibrium EOS to account for the overpressure yielded by the strong electronic heating. The electronic pressure $p_{e}$ depends on the electronic temperature $T_{e}$ according to the Fermi-Dirac model and is expressed as :
\begin{eqnarray}\label{pe}
p_{e}= \frac{2}{3} A \displaystyle\int\nolimits_{0}^{\infty}
\frac{ {\cal{E} }^{3/2}}{1+\exp{\left[({\cal{E}} -\mu)/k_{B}T_{e}\right]}}d{\cal{E}},
\end{eqnarray}
where ${\cal{E}}$ is the internal electronic energy and $k_{B}$ is the Boltzmann constant.  The chemical potential, $\mu$, is determined numerically to match the value of the electronic density, $N_{e}$, at a fixed electronic temperature. The $A$ parameter in equation~(\ref{pe}) is defined as :
\begin{eqnarray}
A=\displaystyle\frac{1}{2\pi^{2}}\left(\displaystyle\frac{2m_{e}}{\hbar^{2}}\right)^{3/2},
\end{eqnarray}
where $m_{e}$ is the mass of the free electron. In order to make a more realistic calculation of the electronic pressure, it is more suitable to use band-structure calculations. Due to the complexity of these calculations, they are beyond the scope of this work. Notwithstanding deficiencies of the Fermi-Dirac model, our simplified approximations are reasonable and give a qualitative view on the behavior of the non-equilibrium system. At solid density, Fermi-Dirac model predicts an electronic pressure contribution in the range of a few hundreds GPa for a temperature of 0K. This degeneracy pressure arises from the Pauli exclusion principle and our model assumes that the electronic contribution only considers the thermal excitation of the free electrons. To improve the description, strong corrections due to screening and chemical bonding must be added and a cold pressure term is often supplied in addition to the electronic and ionic contributions.\cite{Zeldovich} To replace this usual correction, the equilibrium value of the pressure resulting from all contributions has been kept as a reference, and we have subtracted the electronic pressure calculated at ionic temperature and added a value which depends on the electronic temperature. As a consequence, the total pressure used in the above calculations is determined as $p=p_{e}(T_{e})-p_{e}(T_{i})+p_{a}(T_{i})$ where $p_{e}$ and $p_{a}$ are the electronic and atomic pressure, respectively. Note that the atomic pressure $p_{a}$ is given by the employed EOS at the equilibrium temperature and is composed of the electronic, ionic, and cold pressure components : $p_{a}=p_{e}+p_{i}+p_{c}$.

\begin{figure}[htbp]
\includegraphics[width=8.5cm]{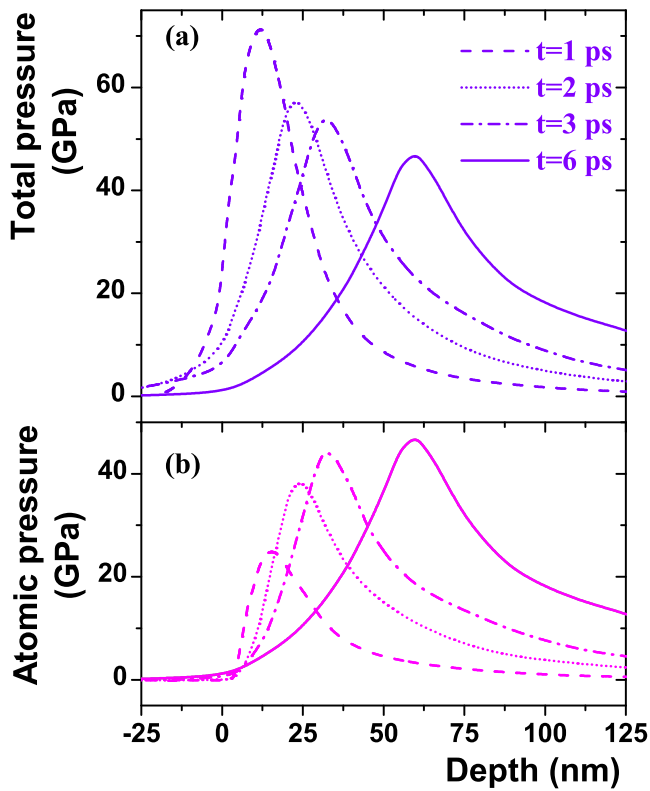}
\caption{(Color online). Non-equilibrium spatial distribution and time behavior of pressure for an aluminum target irradiated by a 150~fs, 5~J cm$^{-2}$ laser pulse indicating the evolution of (a) total pressure and (b) atomic pressure. The atomic contribution is separated from total pressure before complete relaxation.\label{Pr+Prelec}}
\end{figure}

Fig.~\ref{Pr+Prelec} shows the time evolution of the atomic and total pressure profiles induced by a 5~J cm$^{-2}$, 150~fs laser pulse in an aluminum sample. The atomic and total pressure curves are identical at 6~ps due to the fact that electronic contribution vanishes when the relaxation process ends. At this time the maximum amplitude of the atomic pressure is reached. For short times the curves are clearly distinct. This is due to the role played by the electronic pressure in a non-equilibrium state. At this stage, the electronic pressure spreads at a velocity depending on the electronic conduction, which, in turn, is governed in our simulations by the TTM. One can note that the maximum values reached by the partial and the total pressure do not necessarily occur at the same depth, because electronic conduction offsets the highest internal pressure relative to the ionic component. Information about the degree of non-equilibrium can be derived from the difference between electronic and atomic pressure curves. Obviously, the gap between electronic and ionic contribution is more important at shortest times following absorption and close to the surface, where the Fermi gas was subjected to the electromagnetic field. This gap is progressively reduced during the electronic cooling and ionic heating controlled by the TTM, ensuring at the same time energy conservation during heat diffusion and coupling processes. The non-equilibrium remains observable on the first hundred of nanometers, and is especially significant up to 50~nm. The thickness of material affected by strong non-equilibrium pressure is larger than the optical skin depth, which is about 8 nm for aluminum at 800~nm, because the electronic diffusion pushes the electronic pressure towards the bulk material. According to the present EOS, the velocity of the total pressure wave composed by electronic and ionic terms is limited by the sound velocity in near-solid-density aluminum. No shock process due to electronic diffusion velocity exceeding the sound velocity is expected to occur in the simulation. From Fig.~\ref{Pr+Prelec}, decoupling the pressure distribution conveys information about the space and time scales affected by the non-equilibrium dynamics. It results that, at this laser energy density, these non-equilibrium dynamics survive for several picoseconds and can propagate several tens of nanometers under the surface. We will examine in the next section the role of the pressure decoupling of the material properties inside the metal.

\section{shock released thermodynamic states}

In our fluid treatment, internal properties and dynamics of the material evolve jointly and the ablated material is not discernible from the rest of the solid. The use of an ablation criterion is necessary but a definition based on nucleation can be questionable in a one-dimensional simulation. Such models of matter separation of the fluid and solid phases may be responsible for void formation and this feature was not implemented in our calculations. Consequently, the distinction between the ablated and non-ablated material was chosen to be discussed by means of the phase diagram analysis for the irradiated metal, as it will be developed below. To find a picture equivalent of the transformation phenomena that leads to ablation, some authors have previously proposed to study the temporal evolution of the thermodynamic state of the system in search for critical states.\cite{Lorazo03,Perez02,Vidal01} To do this, they discussed phase-space trajectories followed by infinitesimal layers of matter after irradiation in the \emph{(T,$\rho$)} plane of the phase diagram. If several characteristic paths have been identified which may result in material removal, the role of the pressure and its influence on the non-equilibrium has not been indicated, due to the choice of coordinates.

In the following, we consider again the case of a 5 J~cm$^{-2}$, 150~fs laser pulse irradiating metal samples and we will try to connect the space and time observations to pressure-related coordinates for showing the effects of the pressure. We present the successive thermodynamic states in a pressure-density \emph{(p,$\rho$)} phase diagram. The choice of these thermodynamical parameters is well-suited to represent the achievement of new states. Indeed, the electronic contribution is directly visible due to the additional features of the partial pressures. We remind that the hydrodynamical approach used here is based on a Lagrangian formalism and we are studying the behavior of several cells, initially localised at a fixed depth inside the cold homogeneous material preceding the laser pulse arrival. Each cell follows, independently of the others, a thermodynamic path governed by the EOS, providing the pressure as a function of the internal energy and the metal density. Phase-diagrams were obtained by tracking down the pressure time evolution and correlating it with the density behavior based on the EOS. At the considered fluence and due to the limited energy diffusion, only cells located initially in the first hundreds of nanometers show dramatic changes. However, a region of 50 $\mu$m thickness was simulated to avoid spatial confinement of the diffused energy and waves reflection at the rear side of the sample which can happen in thin foils. Simulations have been performed for an ultrashort pulse irradiating two types of metals, aluminum and copper, which differ significantly with respect to their electron-lattice coupling (strong coupling for aluminum, and, respectively, weak coupling for copper). The transport and thermodynamical data for these metals are quite accurately known for equilibrium conditions and this data has been extended to non-equilibrium situations using the Two-Temperature Model framework.\cite{Touloukian,Ebeling,Colombier05} For each metal the set of EOS data fully assigns thermodynamic values over the entire diagram region of phase transitions owing to analytical expressions including specific coefficients which are based on the available experimental and theoretical data. Asymptotic values of the thermodynamical, optical, and transport parameters were fixed to fulfill the equilibrium conditions. This choice ensures a description of the material properties with a good accuracy in the region where thermodynamic constants are accessible and constrains the unknown properties to be fitted with known values. The systematic use of the defined constants at equilibrium for non-equilibrium coefficients limits the number of free parameters and ensures that the specificities of each metal is preserved in the simulation, in an optimal manner with respect to the accessible data.

\begin{figure}[htbp]
\includegraphics[width=8.5cm]{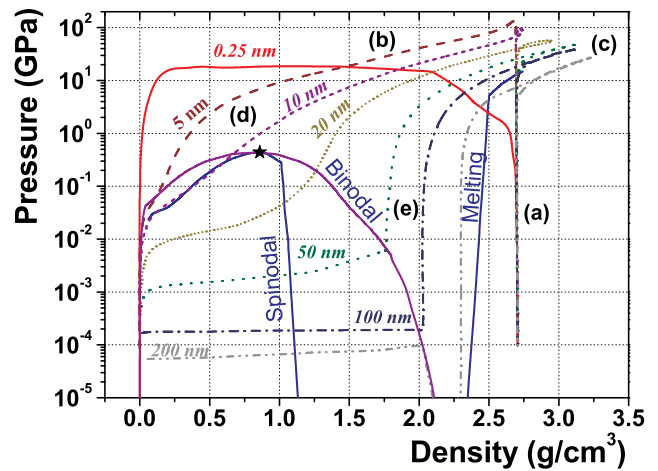}
\caption{(Color online). Pressure-density \emph{(p-$\rho$)} phase diagram of aluminum exposed to ultrafast laser radiation (150 fs, 5 J cm$^{-2}$). Thermodynamic paths of material layers are plotted for several depths inside the metal. The initial depth of the cell is indicated on the top of each curve. The critical point ($\bigstar$), binodal, spinodal, and melting curves are also represented in the diagram.\label{Pro_alu}}
\end{figure}

Fig.~\ref{Pro_alu} depicts the phase diagram of aluminium in the variables pressure and density. The local variations of the matter states during compression and expansion are represented by several thermodynamic paths corresponding to specific locations under the surface of the sample. \emph{(p,$\rho$)} points belonging to a curve describe different time moments of the evolution and no temporal correlation can be easily made between the curves.  Consequently, no relation in time can be directly made between the different paths and only the correlated evolution of the pressure and density remains accessible. However, similar developments occurring in well-specified pressure-density ranges take place on similar time scales. These characteristic behaviors are indicating by labels [$(a)$-$(e)$] on the figure and will be detailed step by step in the following. Consequently, one can roughly estimate the time required for the transformation, as well as the sequence of transitions. The mesh calculation employed here is based on a geometric progression to assure a more precise spatial description in the front of the sample, where gradients are more important and accurate resolution is necessary to sample the skin depth. The thickness of the first cell is fixed at 5 \AA~and the ratio of each two consecutive cell sizes is calculated in order to have the sum of all widths equals to 10 $\mu$m. Seven curves where the mid-points correspond to different approximative depths in the cold solid are represented : 0.25~nm, 5~nm, 10~nm, 20~nm, 50~nm, 100~nm and 200~nm. The numerical convergence has been verified in reducing 5 times the grid cell, leading to less than 2\% change in density and pressure. The transformation phase lines given by the EOS, solid-liquid (melting line) and liquid-gas (binodal line), are added in grey lines on the figure. The spinodal line is also represented to distinguish fluid metastable states to unstable ones. This limit is a theoretical one which marks the location of the infinite compressibility states.\cite{Landau}

The metal starts from solid density (2.7~g cm$^{-3}$ for aluminum) and standard pressure condition (10$^{-4}$~GPa). Then, due to laser heating, the matter expands and passes through different states. During the first picoseconds, electronic heating and the associated non-equilibrium effects generate a strong electronic pressure component. Consequently, a vertical transition in the \emph{(p-$\rho$)} diagram appears, marked by the label $(a)$. The lattice remains cold and a static description of the matter can still be preserved. However, the isochoric heating leads to a sharp increase of the system entropy.  When the laser irradiation stops, the maximum electronic pressure, which corresponds to the peak of compression, has already been reached. The pressure decrease is due to the local volume increase and the cooling of the electron gas by the electron-phonon relaxation $(b)$. After a few tens of picoseconds, the deeper cells enter in an overdense state, in the right part of the thermodynamic diagram $(c)$. The energy transferred from the electron gas to the lattice reduces the non-equilibrium. Additionally, the density decreases due to the progressive ionic heating. At this stage, the thermodynamic trajectories split and follow two paths.

The first one passes over the binodal region because the strong shock release produces a sequence of thermodynamic states above the critical point. The cells located initially at 0.25 nm, 5 nm and 10~nm inside the sample (indicated by the label $(d)$ on Fig.~\ref{Pro_alu}), represent this non-conventional case. The evolution is characterised by a high pressure expansion with no liquid-gas transformation. The expansion of the system takes a few picoseconds during which the fluid reaches a supercritical state and the degree of degeneracy of the electronic subsystem decreases. The swift heating is followed by a fast expansion of the supercritical fluid during this stage. Such thermodynamic states may cause fragmentation, phase explosion, or spinodal decomposition\cite{Perez02} but these effects can not be described in the hydrodynamic frames without involving nucleation theory. Several picoseconds after irradiation, the density drops quickly and the pressure follows a typical ideal gas dependence, evolving according to a $\rho^{5/3}$ law. In such exotic thermodynamic states, the matter exhibits original features and specific plasma effects.\cite{Celliers93,Morikami04,Yoneda03} The second path is followed by the cells originally at 50 nm, 100 nm and 200 nm inside the metal. The associated curves pass through the melting boundary and the liquid maintains a high density while the pressure decreases. Consequently the sound velocity drops and the curves intersect the binodal in the region labeled by $(e)$. The matter enters in the two-phase domain and a liquid-gas transformation occurs. At this stage, which occurs several tens of picosecond after the irradiation, the electron-ion relaxation is complete and the role of electronic pressure becomes negligible. The most important feature in this case is the stepwise decrease of the sound velocity at the point where the quasi-isentropic curve intersects the binodal.\cite{Anisimov02,Sokolowski98} The drastic change in the sound velocity results in a radical structural rearrangement of the flow.\cite{Bulgakova99} In this case, the system relaxes towards the liquid-gas regime within 100-200 ps, and nucleation is supposed to control the main ablation process.\cite{Rethfeld02} The classical theory of homogeneous nucleation determines the formation kinetics of a critical nucleus, depending on the thermodynamics and the transport properties of the fluid. Consequently, the ablation rate for material removed as liquid droplets should present a threshold effect. No consideration of the nucleus size has been accounted for in our 1D calculations and the nucleation process is considered as a homogeneous one.

The limitations to the model impose a series of considerations concerning the material removal process. The amount of ablated material is supposed to be composed of that quantity of matter which attains a supercritical state or undergoes the liquid-gas transition with a positive velocity, seen as limiting cases for material removal. This gives the dimension corresponding to the ablated thickness that we have estimated in a previous study.\cite{Colombier05} The ablated layers that we have defined were composed of the expanded gas and liquid accelerated outside the sample. This definition corresponds to the matter (mixed liquid and gas) reaching a low density, below the critical one. To explore the role of electron-phonon coupling factor, we have performed a set of simulations with a variation of the aluminum coupling constant around the tabulated value. This parameter is characteristic of the TTM and governs the temperature dependence of the energy transfer rate between electron and ion subsystems. It appears that increasing the coupling factor augments the thickness of the material which undergoes supercritical state transition while the thickness of matter which suffers liquid-gas transformation remains relatively unchanged at this fluence. In the TTM framework, the electron thermal conduction plays an opposite role to the electron-ion coupling constant in the control of the energy distribution. In fact, the thermal diffusion parameter allows the propagation of the energy before the complete relaxation. Decreasing the value of the electronic conduction tends to confine the energy to the metal surface and the electronic pressure decreases in a less significant manner due to the cooling associated to the diffusion process. In this case, the pressure remains high and the amount of material reaching supercritical states is more important. However, in order to reach these states, the dynamics of energy transfer is becoming equally important as the shock release, because the unloading wave causes a strong pressure decrease in the front-surface layer. The acceleration of the surface and the associated shock formation are determined by the pressure gradient which, in turn, is governed by the energy profile since in the employed EOS, the pressure is a function of the internal energy and the material density. Due to the hydrodynamic equations, the rapid expansion required for reaching the density of the critical point is driven by a steep pressure gradient. Simulations performed in the case of a weaker thermal conductivity show that the density decreases quickly and the pressure drops jointly. Therefore, the balance between these three competitive processes (energy transfer, thermal diffusion and shock release) determines the rate of achieving supercritical states by the material during its transformation.

\begin{figure}[htbp]
\includegraphics[width=8.5cm]{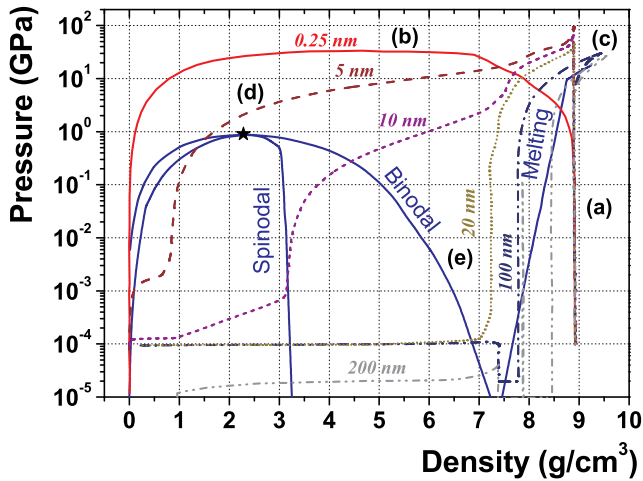}
\caption{(Color online). Pressure-density \emph{(p-$\rho$)} phase diagram of copper  exposed to ultrafast laser radiation (150 fs, 5 J cm$^{-2}$). Thermodynamic paths of specific material layers are plotted for several depths inside the metal. The initial depth of the cell is indicated on the top of each curve. The critical point ($\bigstar$), binodal, spinodal, and melting curves are also represented in the diagram.\label{Pro_cu}}
\end{figure}

To avoid discussions on the parametric values which can be strongly arbitrary, we have chosen to present our results for a second material, copper, which differs strongly from aluminium in that concerns the strength of the electron-phonon coupling. We have preferred this choice of materials with different properties since it can be verified against experimental results. One can note that the experimental and calculated amount of ablated matter is similar for both materials at 5 J cm$^{-2}$, and the thermodynamic states reached by these materials can be discussed in parallel for this fluence used in our simulation. Although these metals have different EOS, the energy deposition process can be compared since it depends mainly on their transport properties. In fact, the critical point is about $\rho_{0}/4$ in density for both materials and the value of the critical pressure does not play a significant role. On the other hand, copper, as mentioned, is characterized by a lower value of the coupling constant. Consequently, longer non-equilibrium processes take place due to a larger thermalisation time. Fig.~\ref{Pro_cu} shows a pressure-density diagram for copper which is similar to the one obtained for aluminum. The diagram depicts thermodynamic paths of the cells at the same depths as previously presented. The curve at 50~nm was skipped because cells located deeper than 20~nm exhibit similar behaviors. The cells start from the solid density at 8.9 g cm$^{-3}$ in this case and follow a short phase of compression at high pressure before the expansion phase. Qualitatively, the liquid-gas transformation resulting from the irradiation of the copper sample shows the same types of paths, corresponding to the expansion of a supercritical fluid layer at the surface of the material. As for aluminum, nucleation may be a possible mechanism of ejection for material situated in the deeper region, up to 200 nm depth. For identical input laser fluence, at 5 J cm$^{-2}$, the supercritical state affects a thicker layer below the aluminum surface than in the copper case due to a stronger absorption. Cells deeper than 5~nm do not present supercritical behavior in the second case while a 15~nm region is becoming supercritical in the first case. Yet, the maximum pressure reached by the outer cell at 0.25~nm is higher for copper. Note that the electronic thermal conduction parameter of copper does not differ significantly from aluminum. However, the electron-phonon coupling is weaker, approximatively $3\times10^{16}$ WK$^{-1} $m$^{-3}$ for copper~\cite{Corkum88} compared to $3\times10^{17}$ WK$^{-1}$m$^{-3}$ for aluminum,\cite{Fisher01} thus the electronic energy is slowly transfered to the crystal lattice allowing for even more diffusion before the energy is relaxed.\cite{coupling} As a consequence, the electronic temperature and pressure are enhanced inside the material in this case. The density decrease, related to the ionic expansion, depends strongly on the energy transfer time. Consequently, the non-equilibrium which survives for a longer time in copper is an impediment for reaching supercritical states because the electronic energy has more time to spread in the metal due to electronic diffusion effects. When the density reaches the region corresponding to the critical point, the majority of the cells have a total pressure lower than the critical pressure and the nucleation process is supposed to take place on a volume larger than the one where fragmentation may occur.

To conclude, it has been shown that the supercritical states are mainly observed at the beginning of the non-equilibrium, for a strong expansion. An interplay between the electromagnetic absorption profile, the absorption coefficient, the electron-lattice coupling, and the electronic thermal conductivity determine the extent of the non-equilibrium. In addition, the shock release, governed by the hydrodynamics and the EOS in our calculation, is responsible for the drop of the pressure at a high or a low density. Expansion of the supercritical states and the slower liquid-gas transformation are identified to be the main regimes of ablation. We have shown that these regimes take place on different time scales. An eventual fragmentation process or phase explosion may occur at the surface several picoseconds after the irradiation. Subsequently, nucleation process is expected to occur within several hundreds of picoseconds. The relative occurence of these ablation processes depends on the metal transport properties which determine the rate of supercritical states reached during the material expansion. The metals considered here show different features mainly due to the strong electron-phonon coupling disparity which exhibits an order of magnitude variation. Since electron-phonon coupling in aluminum is stronger than copper value, and the thermal diffusivities are comparable, aluminum is favored for reaching supercritical states.

\section{Liquid layer : numerical estimation of the size of resolidified material}
After the quasi-isochoric heating and after several intermediate stages including the subsequent plasma formation at the surface, a thin layer confined between the solid-liquid and the liquid-gas interfaces is strongly compressed, forming a hot liquid. If the outer part of this liquid expands quickly, the inner part expands after several picoseconds with respect to the onset of expansion. The expanding plasma and the external liquid layer heat and compress the material for an extended period, much longer than the laser pulse. As a consequence, the melting front propagates further into the material and the liquid thickness increases. Moreover, this compression phase leads to additional material removal via nucleation process. The appearance of Newton fringes on the expanding material, found experimentally using time-resolved imaging of the irradiated regions on short time scales in a large variety of materials, represents clear evidence on the existence of two reflecting surfaces.\cite{Sokolowski98} Visible light reflected by the liquid-gas layer interfering with the retarded light reflected from the dense liquid would explain why the far-field pattern exhibits the characteristic ring structure. This fringes last for several ns and the evaluation of the interference pattern suggest a mixed phase characterized by a high index of refraction. The lifetime of the observed optical interference patterns is a consequence of the ablation of the metal following nucleation. Otherwise, the liquid layer at constant density is not ablated and remains attached to the solid. During this time, thermal conduction cools down the disordered matter which no longer conserves fluid properties characteristic to the liquid state and droplet ejection stops. The long-living liquid layer cannot escape from the sample and resolidification occurs. Significant change in the material structure can occur during the solid-liquid and liquid-solid transformations with the potential of producing new structures with intrinsic properties that are perhaps different from the original ones. The aim of this section is to provide a numerical measure of the thickness of matter concerned by the resolidification phenomenon.

Under ultrafast laser irradiation conditions, the ablation process is highly dependent on the material nature. In dielectric materials, damage and ablation mechanisms are attributed to optical breakdown and plasma formation. For semiconductors, laser irradiation induces electronic transitions from bonding to antibonding states. The presence of a large concentration of excited electrons changes the interatomic forces and the atomic motions at low ionic temperature may be sufficient to break interatomic bonds.\cite{Silvestrelli96,Sundaram02} For metals, the strong electronic excitation does not affect the band structure and the thermal motion represents the most reasonable mechanism for yielding a modification of the atomic configuration which is able to produce deviations from the ordered phase. Large amplitude vibrations result in the loss of the long range order present in the crystalline phase. The structure evolves into a disordered liquid state where only short range atomic correlations are present. Time-resolved electron diffraction experiments have shown that complete solid-to-liquid transition occurs within a few picoseconds for aluminum thin films.\cite{Siwick03} It appears that classical thermal melting features are preserved and equilibrium conditions of melting are supposed to be valid for metals in conditions characteristic to our simulations.

\begin{figure}[htbp]
\includegraphics[width=8.5cm]{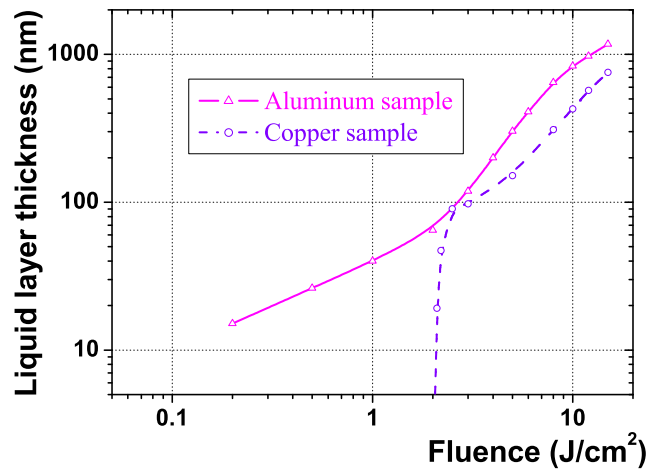}
\caption{(Color online). The thickness of the liquid layer remaining on the metal substrate at the crater bottom versus the incident laser fluence for aluminum (solid line) and copper (dashed line) samples. Irradiation conditions: 150 fs laser pulse. \label{Liquid}}
\end{figure}

To determine whether a numerical layer undergoes a solid-to-liquid transformation or not, we have examined the specific internal energy value $E$.  We compare this value with the melting energy $E_{m}$, whose values in our simulations are the ones corresponding to the plateau onset provided by the EOS. For aluminum at Standard Temperature and Pressure conditions (STP), $E_{m}^{STP}=6.74\times10^{5}$ J kg$^{-1}$, and this value grows significantly as the pressure increases. For copper, the dependence is less significant than for aluminum in the considered pressure range and the energy shows a reduced scattering around the standard value $E_{m}^{STP}=4.7\times10^{5}$ J kg$^{-1}$. When the local energy is sufficiently high, a solid-to-liquid transition starts and crystal modifications should occur. In this study, we only consider the non-expanding material and the thickness of the liquid layer is calculated by determining both the positions of the outer cell which undergoes liquid-gas transformation and the first one from the bulk that has an energy $E\geq E_{m}$. The difference between these two locations of the cells provides the size of the liquid confined between the solid substrate and the expanding ablated matter. Fig.~\ref{Liquid} shows the liquid layer thickness as a function of the laser fluence for aluminum (solid line) and copper (dashed line). It demonstrates the pronounced dependence of the liquid-gas transformation and heat diffusion effect on the incident laser fluence.

While liquid droplets generation requires strict kinetics and thermodynamic conditions, no restrictions apply concerning the liquid layer formation mechanism. Examination of Fig.~\ref{Liquid} and previously performed simulations reported in Ref.~[\onlinecite{Colombier05}] reveal that for fluence higher than 5 J cm$^{-2}$ the thicknesses of the liquid layer and, respectively, the ablated matter are of the same order of magnitude for both materials. For low fluences, a large part of the heated aluminum sample is ablated and a small quantity of liquid remains on the solid substrate. As a consequence, only tens of nanometers of liquid layer result from the irradiation just above the threshold fluence for aluminum ablation. This is in agreement with  experimental observations of a limited recast material at the threshold fluence. For copper, the calculated liquid thickness is zero at low fluences, below the melting threshold, and increases quickly close to the threshold fluence, $F_{Th}^{Cu}$. The value of $F_{Th}^{Cu}$ is not realistic and the discrepancy has been already discussed in Ref.~[\onlinecite{Colombier05}], but the curve in Fig.~\ref{Liquid} shows a correct dynamic behavior. The discrepancy is apparently due to the effects of the non-equilibrium transport on the optical properties, which was not accurately described in our simulation. The slope of the curve for copper is more pronounced than for aluminum due to the higher electronic heat diffusion characteristic for copper that spreads the energy in a larger volume. Thus, only fluences very close to $F_{Th}^{Cu}$ allow to reach a minimal size of recast material. Finally, even if $E_{m}$ is higher for aluminum, copper is denser than aluminum and its melting temperature is 1.5 times higher. The local energy required for melting aluminum is lower and simulations show a thicker liquid layer for this material.

To control the quality of the laser micro-machining process, the region undergoing mesoscopic irreversible transformation has to be reduced. Then, due to the correlation between the resolidification size and the laser fluence, it is important to connect macroscopically-induced effects to the laser parameters. The numerically-estimated liquid size can be directly compared to the recast size in the hypothesis of a one-dimensional model. These results emphasize the interest in using the threshold fluence to minimize recast material and improve the ablation quality. Appropriate irradiation processes may be used to create very high quality micro-machined structures and simulations are able to predict the optimal intensity range.

\section{Spallation regime of ablation}
To add to the description of the potential factors that may end up in ablation upon ultrafast laser irradiation, this section is devoted to laser-assisted spallation process taking place in the solid phase. The formation and propagation of shock waves resulting from isochoric heating can lead to high tensile stress close to the rear surface of the sample if it is thin enough. Consequently, spallation follows. The peak of the pressure wave results in a shock front, while its rear decreasing slope induces a rarefaction wave into the shocked material. When the shock wave reaches the free surface, the compressed material expands freely and accelerates the surface expansion. The superposition  of the two rarefaction waves, the tail of the initial pressure wave and tensile stress wave, running in opposite directions, yields a residual stress whose amplitude increases with the distance from the free surface in the early times. The mechanism of mechanical fracture is due to the nucleation of microvoids which grow in the region where high plastic deformation occurs. As the load on the metal increases, the microvoids eventually form a continuous fracture. In the simulation results reported here, the characteristic of the stress profile induced in aluminum samples is compared with the experimental detection of spallation phenomena performed by Tamura~\emph{et al}~\cite{Tamura01} who measured a linear relationship between foil thickness and the depth where spallation occurred, i.e., spall thickness.

Coupled molecular-dynamics and fluid computational models have revealed high tensile stress leading to the mechanical disintegration of thin films.\cite{Ivanov03b} Vidal \emph{et al.}\cite{Vidal04} have recently performed hydrodynamical simulations to reproduce spallation experiments. It was argued that standard defect growth models are irrelevant for fast hydrodynamic regime and the spall strength used in the above-mentioned simulations is about one order of magnitude higher than the spall strength measured at strain rates in the range ($10^{3}$-$10^{6}$)~s$^{-1}$.\cite{Moshe00} The spall strength value used in the respective study corresponds to the crossing-point between spinodal and cold pressure curves given by the Quotidian-Equation-Of-State (QEOS) model.\cite{More88} This EOS displays a van der Waals behavior which allows tensile stress calculations but does not take into account defects present in solids. However, the mechanical properties of solids are controlled by lattice imperfections and dislocations, which weakens the materials. This deviation from a perfect lattice may explain the large discrepancy between experimental and computational conditions of irradiation proposed by these authors in their simulations and the very high incident fluence levels that were used. The aim of this section is to provide new mechanical insights into the spallation process, while preserving the experimental conditions usual for ablation experiments. Additionally, we will determine and discuss the spall strength value.

\begin{figure}[htbp]
\includegraphics[width=8.5cm]{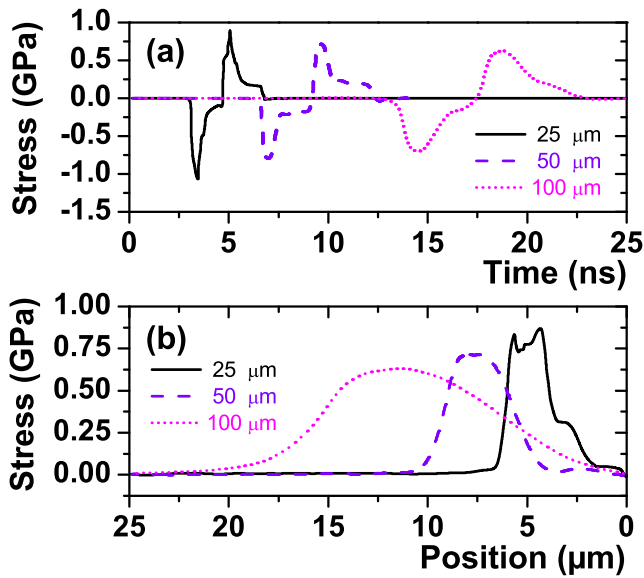}
\caption{(Color online). (a) Stress evolutions in time at the position of the maximum tensile stress for three different numerical simulations performed in aluminum target of 25 $\mu$m, 50 $\mu$m, and 100 $\mu$m thicknesses. (b) Stress profiles at the time corresponding to the maximum tensile stress. Parameters used in the simulations are chosen to reproduce irradiation conditions of Ref.~\onlinecite{Tamura01} : 25 $\mu$m thickness, 25 J cm$^{-2}$ pulse (solid line), 50 $\mu$m thickness, 35 J cm$^{-2}$ pulse (dashed line), and 100~$\mu$m thickness, 70~J cm$^{-2}$ pulse (dotted line).\label{spall}}
\end{figure}

The measurements of the spall thickness reported by Tamura \emph{et al.}\cite{Tamura01} are of interest because they were obtained using irradiation levels close to the mechanical ablation threshold. From this point of view we can assume that lower fluences and the associated tensile stress are not sufficient to produce fractures. Consequently, at the mechanical ablation fluence threshold, neglecting dynamical effects, the position corresponding to the maximum amplitude of the tensile stress can be directly compared to the spall thickness and supplies the criterion for the determination of the mechanical strength. We have performed simulations of the effects induced by a $\tau=50$ fs (FWHM) laser pulse at a wavelength of 775~nm to compare the calculated stress evolution to the experimental measurement of the fracture position. The $s$-polarised incident laser irradiates the target at 57.5\degre with respect to the surface normal. Note that the state of laser polarisation was not specified in the reported experiment and different polarisation cases have been numerically tested but none of them modified the results significantly. The hydrodynamic responses of three targets of different thicknesses: 25 $\mu$m, 50 $\mu$m, and 100 $\mu$m, were studied under laser intensities of 0.5$\times 10^{15}$, 0.7$\times 10^{15}$, and 1.4$\times 10^{15}$ W cm$^{-2}$ respectively. Fig.~\ref{spall} shows the time evolution of the stress at the location of the maximum tensile stress (a) and the stress profiles at the time of maximum tensile stress from the rear surfaces of the targets (b).

Our main concern in these simulations was to follow the position of the maximum tensile stress and to deduce the value which generates a void in the sample. For the 25~$\mu$m thick foil, numerical considerations indicate a maximum tensile stress and, therefore, a probable fracture at 4.4~$\mu$m depth from the rear face of the target. For thicker samples, the spall thickness increases and is estimated to be 7.2 $\mu$m and 11.3 $\mu$m for 50 and 100 $\mu$m foil thickness, respectively. The numerically-determined values and the experimental values of the spall thickness are reported in table~[\ref{Spalltab}]. Maximum tensile stress is also indicated and this provides an estimation of the static tensile strength related to a high strain rate. Good agreement is obtained between the experimental and the simulated spall thicknesses which depend mainly on the velocity of the rarefaction wave front. Details and reasoning about the modeling approach are given below. Following the usual approach in stress-related problems, the stress is decomposed into a volumetric and a deviatoric part. The volumetric part of the stress is computed using the EOS, while the deviatoric one is calculated using an additive decomposition of the rate of deformation into elastic and plastic parts, the von Mises yield condition,\cite{vonMises} and a flow stress model based on Steinberg-Cochran-Guinan (SCG) formalism.\cite{Steinberg80} The SCG model is applicable at high strain rates and has been used to provide a temperature and pressure variation of the yield strength and the shear modulus. To correctly describe the material response under shock conditions, material-strength model should be based on the evolution of the micro-structural state which would required the dynamics of dislocations and grain-boundaries.

\begin{table}[htbp]
  \begin{ruledtabular}
  \begin{tabular}{ccccc}
  %\hline\hline
   \multicolumn{1}{c}{Foil Thickness}& \multicolumn{1}{c}{$F_{Th}^{S}$} &\multicolumn{2}{c}{Spall Thickness}&\multicolumn{1}{c}{$\sigma_{S}$}\\
   %& Spall Thickness & Spall Thickness & Tensile Strength \\
 %  ($\mu$m)&($J/cm^{2}$) & Simulated ($\mu$m)& Experimental ($\mu$m)& (GPa) \\
  ($\mu$m)&(J cm$^{-2}$) & Sim. ($\mu$m)& Exp. ($\mu$m)& (GPa) \\

  \hline
  25 & 25 &$4.4$ &$2.9 \pm 0.9$ &$0.88$ \\
  50 & 35 &$7.2$ &$6.3 \pm 1$ &$0.72$\\
  100 & 70 &$11.3$ &$11.2 \pm 1.2$ &$0.62$ \\
  \end{tabular}
\end{ruledtabular}
\caption{Experimental (Exp.)~\cite{Tamura01} and simulated (Sim.) values of the spall thickness and the associated tensile strength $\sigma_{S}$ given for three different Al foil thicknesses.\label{Spalltab}}
\end{table}

It is worth noting that the mechanical model used in this work is an elastic-perfectly plastic model involving stress behavior which responds elastically up to the point where the stress exceeds the yield stress and after which plastic flow occurs in the material. When the solid undergoes plastic deformation at high pressure, stresses result in the generation and subsequent propagation of lattice dislocations. Whereas solid state plastic flow is caused by the rearrangement of the atoms in the crystalline matrix by transporting these dislocations, such an empirical model does not consider changes in lattice structure. The density of the generated dislocations and their propagation velocity are both dependent on the rate of the applied strain.\cite{Kalantar00} The parameters defined in the constitutive models are determined from experiments involving strain rates on the order of $10^{3}-10^{6}$ s$^{-1}$. Application of the model is speculative at higher strain rates of $10^{7}-10^{9}$ s$^{-1}$ and should be compared carefully to experimental data. In practice this study is difficult and the numerical representation of the stress flow is obtained by estimating the physical parameters involved under deformations. Such parameters are calibrated in experiments involving strain rates on the order of $10^{3}-10^{6}$ s$^{-1}$ and the applicability of the SCG model for higher strain rate is questionable. Nevertheless in our simulations, high strain rates on the order of $10^{9}$ s$^{-1}$ are reached at the front side of the sample and, at the same time, high volumetric stress levels provided by the EOS are attained. In this case, the deviatoric contribution does not play a significant role compared to the volumetric one. As a consequence, the effects of material strength may be ignored and the hydrodynamic response governs the mechanical deformation of the irradiated surface. In spite of uncertainties, we assume that the accuracy provided by the SCG model is acceptable to take into account the pressure dependence of the shear modulus and the yield strength in this hydrodynamic regime. On the other hand, after such wave propagation, the stress is much lower than at the front side of the sample and the elasto-plastic properties of the material play a significant role for pressures below 1~GPa. After the shock has propagated several micrometers, the stain rates decrease below $10^{6}$ s$^{-1}$ and we suppose that the shock response of the metal can still be modelled by the parameters associated to the SCG model. In fact, this shock compression at lower strain rates belongs in the validity domain of the SCG constitutive model and corresponds to a regime where solid state material strength is strongly affected by the enhancement of pressure and the temperature. Calculated values of the tensile strength are lower than the standard value (rate-independent) around 1 GPa for lowest strain rates. However, high strain rate spallation experiments show that for increasing loading rates in the tensile region before spallation, the spall strength increases with the rate of mechanical loading and reach several GPa for strain rates around 10$^{7}$ s$^{-1}$.~\cite{Moshe98} Consequently, the deduced value of the spall strength is lower than expected, which may reveal a limitation in the mechanical model under such strength of material deformation. Local defects may also affect the experimental threshold fluence of spallation. The observed decrease of the calculated spall strength while the foil thickness increases may be due to the strain rate decrease with the propagation time of the shock wave. Nevertheless, the spall strength values determined by the simulation are close to the ones corresponding to the low strain rate range.

\section{Conclusion}
We have presented a one-dimensional numerical tool that yields information on several aspects of ultrafast laser metal ablation. The insertion of non-equilibrium aspects into a Lagrangian hydrodynamic code describing laser-matter interaction has provided a novel ultrafast hydrodynamic perspective of matter ejection. The distinction between the time evolution of the atomic pressure and the total pressure has revealed a strong electronic contribution on the first tens of nanometers for several picoseconds after irradiation. Due to this electron-ion non-equilibrium, unconventional thermodynamic states can be reached and we proposed a synthetic representation of the thermodynamic matter evolution in suitable diagrams for copper and aluminum in the view of their different efficiencies in the electron-phonon coupling. Then, the successive thermodynamic properties were plotted at different depths inside the metal to show layers trajectories in a \emph{(p-$\rho$)} phase-plane. The fast laser absorption at the metal surface yields isochoric heating and the solid was brought in states far from equilibrium. The relaxation process occurs through the expansion phase and we have identified different paths corresponding to different mechanisms for matter ejection. The expansion of the supercritical fluid, which may result in a fragmentation process, and a potential homogeneous nucleation mechanism have been identified as possible ablation processes. It has been shown that the supercritical states responsible for the fragmentation process are mainly observed in the first picoseconds of non-equilibrium, owing to a strong expansion. We emphasize that the materials with a high electron-phonon coupling are then favored in reaching such states. Further informations on the phase transition under such irradiation condition may be obtained by combining molecular-dynamic description to fluid treatment to avoid assumptions inherent to the use of EOS.

An estimation of the thickness of the liquid layer, localised between the expanding matter and the unaffected solid substrate, was performed. It can be related to the ablation rate and structure size in the longitudinal direction and could be compared with post-mortem examination. The control of the quantity of recast material would be important to improve micro-structuring quality for micro-machining applications. Mechanical fracture of bulk aluminum was also investigated. If experimental spall thicknesses were correctly reproduced, our deduced values of the spall strength are on the same order of magnitude as the ones corresponding to the low strain rate range. These values are far below the values measured in dedicated experiments and suitable void nucleation models are needed to correctly describe solid fragmentation. However, our simulations provide reliable information on both thermodynamic and mechanic properties under such extreme conditions.

\section*{ACKNOWLEDGEMENTS}
The authors are grateful to N. Bulgakova and F. Bonneau for fruitful discussions on several aspects of this work.

\end{document}